\documentclass[12pt]{article}
\usepackage{amssymb,amsmath}
\begin{document}


\sloppy
\title
{\Large  Classical and quantum scattering by a Coulomb potential }

\author
 {
       A.I.Nikishov
          \thanks
             {E-mail: nikishov@lpi.ru}
  \\
               {\small \phantom{uuu}}
  \\
           {\it {\small} I.E.Tamm Department of Theoretical Physics,}
  \\
               {\it {\small} P.N.Lebedev Physical Institute, Moscow, Russia}
  \\
 }
%
\maketitle
\begin{abstract}
For relativistic energies the small angle classical cross section for
 scattering on a Coulomb potential agrees with the first Born approximation
 for quantum cross section for scalar particle only in the leading term. 
 The disagreement in other terms can be avoided if the sum of all corrections 
 to the first Born approximation for large enough Coulomb charge contain the
 classical terms which are independent of that charge. A small part of the
 difference in classical and quantum cross sections may be attributed to the 
 fact that the relativistic quantum particle can rush through the field without
 interaction. We expect that smaller impact parameters and spin facilitate 
 this affect.
\end{abstract}   
\section{Introduction}
The initial motivation for this consideration was the interest in the 
properties of gravitational field. The heuristic approach to gravity [1-3]
suggests that the Riemannian space appears as a result of changing by the
gravitational field of measuring rods and clocks. If so, the formation
of changes in the rods and clocks requires some time and something new
can occur when the formation time become comparable with the period of
gravitational frequency. In particular, one can imagine the situation when
a relativistic particle can rush through the field so quickly that (with a
non-zero probability) no interaction with the field occur.

Similar event can happen in a Coulomb scattering. In the nonrelativistic
case the quantum particle scattering is governed by the same Rutherford 
formula as the classical one. In classical picture each particle is scattered
by the field, so the above mentioned agreement should mean that passing through
the field without interaction is impossible.
In the relativistic region classical and quantum cross sections are different
and a part of the difference may be attributed to the fact that quantum 
particle can fly through the field without being deflected. The classical 
cross section in small angle region have terms which have different signs
 for attractive and repulsive potentials. Their analogue in quantum case 
 contain the Planck constant $\hbar$ in the denominator of the fine structure
  constant $\alpha=\frac{e^2}{\hbar c}$.
 This suggest that unless $\alpha Z$ is made sufficiently 
 large the reproduction of classical terms from quantum ones is impossible.
  For this reason the theoretical investigation of the
  relationship between the classical and quantum scattering is of interest. 
  It is even probable that some cleverly designed experiments can also clarify
   the situation.

\section{Scattering by attractive potential}

Now we are going to get the classical cross section. 
The trajectory of a classical particle in Coulomb field was obtained 
by C. G. Darwin in 1913 [4]. The scattering angle $\theta$ as a function of
impact parameter $\rho$ and velocity at infinity $\beta=v/c$ is given 
by the relation
$$
\frac{\theta}{2}=\frac{\pi-\chi}{\sqrt{1-z^{-1}}}-\frac{\pi}{2},  \eqno(1)
$$
where
$$
\chi=\arctan\xi,\quad \xi=\beta\sqrt{z-1}\quad z=
\left(\frac{\rho}{a}\right)^2\frac{\beta^2}{1-\beta^2},\quad a=
\frac{ee'}{mc^2}. \eqno(2)
$$
The integral cross section is
$$
\sigma_{cl}(\theta)=\pi\rho^2(\theta),                              \eqno(3)
$$
because all particles with impact parameters less then $\rho$ are scattered,
see Problem 1 in \S 39 in [5]. We note here that $(ee')^2$ enters into the
classical cross section only as a factor $a^2=(\frac{ee'}{mc^2})^2$. This follows
from (1) and (2) as $\rho/a$ is a function of only $\theta$ and $\beta$.

We expect that classical approach is justified for large angular momentum
$\rho p$ ($p$ is the momentum at infinity) i.e. for large impact parameters.
 So we  may assume that $\frac{\hbar}{\rho p}<<1$ is the accuracy of classical
 treatment. 
We  can write 
$$
\sqrt z=\frac{\rho p}{\hbar}\frac{\hbar c}{ee'}.           \eqno(4)
$$
We note that in the nonrelativistic
region when $c\to\infty$ also $z\to\infty$. As seen from (4) the condition 
$\frac{\rho p}{\hbar}>>1$ can be satisfied even for $z$ of the order unity
if $\frac{ee'}{\hbar c}$ is correspondingly large. 

In general we may write
 $$
  \frac{d\sigma_{cl}}{d\theta}=\pi\frac{1}{|\frac{d\theta}{d\rho^2}|}, \eqno(5)
  $$
  where $\frac{d\theta}{d\rho^2}$ should be calculated from (1) and (2) and
  taken at $\rho^2$ obtained numerically from  (1) for the considered 
  $\theta$.

For $\xi, z>>1$ we have
$$
\chi \equiv \arctan\xi=\frac{\pi}{2}-\frac{1}{\xi}+\frac{1}{3\xi^3}-
\frac{1}{5\xi^5}+\cdots,                                              
$$
$$
\frac{1}{\xi}=\frac{1}{\beta\sqrt z}(1-z^{-1})^{-1/2}=
\frac{1}{\beta}\sqrt{\varepsilon}\{1+\frac12\varepsilon+
\frac{3}{2^3}\varepsilon^2+\cdots\}, \quad \varepsilon=\frac1z.        \eqno(6)
$$
From (6) it follows
$$
\chi=\frac{\pi}{2}-\beta^{-1}\sqrt{\varepsilon}+[-\frac{1}{2\beta}
+\frac{1}{3\beta^3}]\varepsilon^{3/2}+[-\frac{3}{2^3\beta}+
\frac{1}{2\beta^3}-\frac{1}{5\beta^5}]\varepsilon^{5/2}+\cdots . \eqno(7)
$$
Using also
$$
(1-z^{-1})^{-1/2}=\{1+\frac12\varepsilon+
\frac{3}{2^3}\varepsilon^2+\cdots\},                    \eqno(8)
$$               
we find for $\frac{\theta}{2}$ in (1)
$$
\frac{\theta}{2}=\frac{1}{\beta}\sqrt{\varepsilon}+
\frac{1}{2^2}\pi\varepsilon +[\frac{1}{\beta}-
\frac{1}{3\beta^3}]\varepsilon^{3/2}+\frac{3}{2^4}\pi\varepsilon^2+
[\frac{1}{\beta}-\frac{2}{3\beta^3}+\frac{1}{5\beta^5}]\varepsilon^{5/2}+\cdots.
                                                                   \eqno(9)
$$
This gives the scattering angle $\theta$ as a function of impact parameter
$\rho$, see the expression for $z$ in (2), $\varepsilon=z^{-1}$.

 Inverting (9) with the help of eq. (3.6.25) in [6] we obtain
 $$
 z^{-1/2}=\varepsilon^{1/2}=A\frac{\theta}{2}+B\left(\frac{\theta}{2}\right)^2+
 C\left(\frac{\theta}{2}\right)^3+D\left(\frac{\theta}{2}\right)^4
 +E\left(\frac{\theta}{2}\right)^5+\cdots,                          \eqno(10)
 $$
 where                                             
 $$ A=\beta; \quad B=-\frac{\pi}{2^2}\beta^3;\quad C=\frac13\beta-\beta^3
 +\frac{\pi^2}{2^3}\beta^5;\quad D=-\frac{5\pi}{2^2\cdot3}\beta^3+
 \frac{17\pi}{2^4}\beta^5-\frac{5\pi^3}{2^6}\beta^7;
 $$
  $$
  E=\frac{2}{3\cdot5}\beta-\frac{2^2}{3}\beta^3+
  \left(2+\frac{7\pi^2}{2^4}\right)\beta^5-\frac{3\cdot11}{2^5}\pi^2\beta^7+
  \frac{7\pi^4}{2^7}\beta^9.                                        \eqno(11)
  $$

  Next we rewrite (10) in the form
  $$
  z^{-1/2}=\beta\frac{\theta}{2}\{1+a_1\frac{\theta}{2}+
   +a_2\left(\frac{\theta}{2}\right)^2 +a_3\left(\frac{\theta}{2}\right)^3+
   a_4\left(\frac{\theta}{2}\right)^4+\cdots \}          \eqno(12)
  $$
  From (12) with the help of eq. (3.6.17) in [6] we get
  $$
  z=\left(\frac{2}{\beta\theta}\right)^2f(\theta/2,\beta),\quad
  f(\theta/2,\beta)=1+\frac{\pi\beta^2}{2}\frac{\theta}{2}+
  [-\frac23+2\beta^2-\frac{\pi^2\beta^4}{2^4}]\left(\frac{\theta}{2}\right)^2+
  $$
 $$
 +[\frac{\pi\beta^2}{3}-\frac{5\pi\beta^4}{2^3}-
 \frac{\pi^3\beta^6}{2^5}]\left(\frac{\theta}{2}\right)^3+
[\frac{1}{3\cdot5}+\frac{2\beta^2}{3}-
\left(1+\frac{\pi^2}{2^2}\right)\beta^4+\frac{3\cdot5\pi^2\beta^6}{2^5}-
\frac{5\pi^4\beta^8}{2^8}]\left(\frac{\theta}{2}\right)^4+
 \cdots 
    \eqno(13)
  $$
  It follows from (2) that
  $$
  \rho^2=a^2\frac{1-\beta^2}{\beta^2}z.   \eqno(14)
  $$
  Using here (13) we find
  $$
 \sigma_{cl}(\theta)=\pi\rho^2(\theta)=
  \pi\left(\frac{ee'}{vp}\right)^2\frac{1}{(\theta/2)^2}f(\theta/2,\beta);\quad
  p=\frac{mv}{\sqrt{1-\beta^2}},                         \eqno(15)
  $$
  where $f(\theta/2,\beta)$ is given in (13).
  We see that terms of odd powers of $\theta$ are present in classical
 cross section. They have negative sign in the repulsive case, see below.
 This has clear physical explanation: in attractive case the particle comes 
 closer to the center and is scattered in a larger angle. These terms are
 absent in quantum  formula  in Born approximation and the reason is also 
 clear: the particle moves freely before and after a single interaction and 
 the sign of particle's charge is unimportant in the cross section. So, using
 wave packets to fix the impact parameter will not change the situation. 
 The correction to Born approximation do  depend on the sign of particle's
  charge and is proportional to $\frac{ee'}{\hbar c}$ i.e. to $\alpha Z$ and
   Planck
 constant is in the denominator. This suggests that the sum of all corrections
  can contain the classical terms if $\alpha Z$ is sufficiently large. The 
  differential  cross section is obtained from the integral one (15) by
   differentiation over $\theta$ (and changing the overall sign):
 $$
 \frac{d\sigma_{cl}}{d\theta}=
 \pi\left(\frac{ee'}{vp}\right)^2\{\frac{8}{\theta^3}+\frac{\pi\beta^2}
 {\theta^2}+
 [-\frac{\pi}{6}\beta^2+\frac{5}{16}\pi\beta^4+\frac{1}{64}\pi^3\beta^6]+
 $$
 $$
 [-\frac{1}{30}-\frac{1}{3}\beta^2+(\frac12+\frac{1}{8}\pi^2)\beta^4-
 \frac{15}{64}\pi^2\beta^6+\frac{5}{512}\pi^4\beta^8]\theta+\cdots\}. \eqno(16)
 $$
  Now we can compare the classical cross section with the quantum one for a
  scalar particle
  $$
  \sigma_{qu}(\theta)=\pi\left(\frac{ee'}{vp}\right)^2
  \int\limits_{\theta}\limits^{\pi}
  \frac{\cos(\theta/2)}{\sin^3(\theta/2)}d\theta=
  \pi\left(\frac{ee'}{vp}\right)^2\cot^2(\theta/2)=
  $$
  $$
  \pi\left(\frac{ee'}{vp}\right)^2
  \frac{1}{(\theta/2)^2}\{1-\frac23\left(\frac{\theta}{2}\right)^2
  +\frac{1}{3\cdot5}\left(\frac{\theta}{2}\right)^4+\cdots \},     \eqno(17)
  $$
  see, for example [7]. We note that $f(\theta/2,\beta)$  for $\beta\to0$
  goes over to the expression in braces in the right hand side of (17).
 This have to be expected as (17) is essentially the (integral) Rutherford 
 cross section.

 Finally, the coefficient in front of $(\theta/2)^4$ is larger in classical
 cross section than in quantum one. It is enticing to interpret this in 
 such a way the  that the quantum particle can fly through the field without
 interaction. Such a possibility is of great interest for gravitational field
 where the analogue of $ee'$ is $Gmm'$ and is not restricted.

 If the scattered particle have spin $1/2$ the quantum cross section acquires 
 the additional factor $(1-\beta^2\sin(\theta/2))$, see eq. (7.22) in [8].
 It looks like the spin of a particle and smaller impact parameters 
 (corresponding to larger $\theta/$) facilitate the flight without
 interaction.

 Returning to equation (17), we note that the integrand gives the
 differential cross section and
 $$
\frac{\cos(\theta/2)}{\sin^3(\theta/2)}=\frac{8}{\theta^3}-\frac{1}{30}\theta
+\cdots.                                                \eqno(18)
$$
 For the electron we have
 $$
 \frac{\cos(\theta/2)}{\sin^3(\theta/2)}(1-\beta^2\sin^2(\theta/2))=
 \frac{8}{\theta^3}-\frac{2\beta^2}{\theta}+\left(\frac{\beta^2}{6}-
 \frac{1}{30} \right)\theta+\cdots.   \eqno(19)
$$

  \section{Scattering by repulsive potential}

  In this case instead of (1) we have
 $$
\frac{\theta}{2}=\frac{\pi}{2}- \frac{\chi}{\sqrt{1-z^{-1}}},  \eqno(20)
$$
see Problem 1 in \S 39 in [5]. (The angle $\varphi_0$ in [5] is 
the half angle between asymptotes, not the whole angle as misprinted there.)
In the same manner as before we obtain (10) , where $A$, $C$ and $E$ are
the same as in (11), but $B$ and $D$ have opposite sign, i.e. they are
obtained from (11) by substitution: $B\to-B$, $D\to-D$. Correspondingly
in the expression for $z$ in (13) and in (15) we have to make the
 substitution $\theta\to-\theta$ and in (16) also $d\theta\to-d\theta$.

We note also that (3) simply states that  each classical particle with 
the impact parameter $\rho<\rho(\theta)$ is scattered. The differential
cross section is obtained from the integral one in (3) by differentiating
over $\theta$ (and changing the sign). For $\theta$ not small enough the
classical approach is inapplicable, but the integral cross section should
be valid as the "total" integral  cross section (for small  $\theta$) in
 the sense that it include inelastic processes (as bremsstrahlung) and
 absorption by some disk around Coulomb center.

 \section{Conclusions}
 The classical cross section for scattering in relativistic region is 
 different for attractive and repulsive potentials and do not agrees 
 exactly with quantum cross section for scalar particle. It seems that this
 disagreement cannot be totally ascribed to the fault of classical approach.
 This suggest that quantum corrections to the first Born approximation should
 have such a structure that for sufficiently large $\alpha Z$ they give 
 classical terms which are independent of $\alpha Z$.

 It is not excluded that with small probability a high energy particle
 can rush through Coulomb field without deflection.
  \section {Acknowledgements}
 I am greatly indebted to V.I. Ritus for fruitful discussions.

 The work was carried out with financial support  of Scientific Schools
 and  Russian Fund for Fundamental Research  (Grants 4401.2006.2 and 
 05-02-17217).

 \section*{References}
1. Dehnen H., H\"onl H., and Westpfahl K., Ann.  der
          Phys. {\bf6}, 7 Folge, Band 6, Heft 7-8, S. 370-407, (1960).\\
2.   Thirring W.E., Ann. Phys. (N.Y.) {\bf16}, 96-117,(1961).  \\
3. Nikishov A.I. arXiv:0710.4445, v1 [gr-qc] 24 Oct 2007.\\
4. Darwin C.G. Phil. Mag.(6) {\bf 25}, p. 201 (1913).\\ 
5.Landau L.D. add Lifshitz E.M.  {\sl The Classical theory of Fields}, 
Cambridge, MA (1971)\\
6.Abramowitz M. and Stegun I. { \sl Handbook of Mathematical Functions},
National Bureau of Standards (1964).\\
7. Pauli W. Rev. Mod. Phys. {\bf 13}, 203-232 (1941).\\
8. Bjorken J.D. and Drell S.D.{ \sl Relativistic Quantum Mechanics.} McGraw-Hill
Book Company, New York (1964).                                                   
\end{document}